\documentclass[11pt]{report}
\usepackage{anysize}
\usepackage{graphicx}
\usepackage{amsmath}
\usepackage{amssymb}
\usepackage{wrapfig}
\usepackage{boxedminipage}
\usepackage{algorithm}
\usepackage{algorithmic}

\usepackage{framed}
\usepackage{color}
\usepackage{aabrf}

\newenvironment{packed_enum}{
\begin{enumerate}
  \setlength{\itemsep}{1pt}
  \setlength{\parskip}{0pt}
  \setlength{\parsep}{0pt}}{\end{enumerate}
}
\newenvironment{packed_item}{
\begin{itemize}
  \setlength{\itemsep}{1pt}
  \setlength{\parskip}{0pt}
  \setlength{\parsep}{0pt}}{\end{itemize}
}

\title{EMBANKS: Towards Disk Based Algorithms For Keyword-Search In Structured Databases}
\date{}
\author{Nitin Gupta}

\begin{document}
\thispagestyle{empty}

\begin{titlepage}

\begin{center}

\vspace{\baselineskip}
\vspace{\baselineskip}
\vspace{\baselineskip}
\textbf{{\large EMBANKS: Towards Disk Based Algorithms For Keyword-Search In
Structured Databases}}\\ {\large
\vspace{\baselineskip}
\vspace{\baselineskip}
\vspace{\baselineskip}
\vspace{\baselineskip}
\vspace{\baselineskip}
Submitted in partial fulfillment of the requirements\\
for the degree of\\
\vspace{\baselineskip}
\textbf{Bachelor of Technology}\\
\vspace{\baselineskip}
\textbf{Nitin Gupta}\\
\vspace{\baselineskip}
\vspace{\baselineskip}
\vspace{\baselineskip}
Advised by\\
\textbf{Prof. S. Sudarshan}\\
\vspace{\baselineskip}
\vspace{\baselineskip}
\vspace{\baselineskip}
\vspace{\baselineskip}

\begin{figure}[htbp]
\begin{center}
\includegraphics{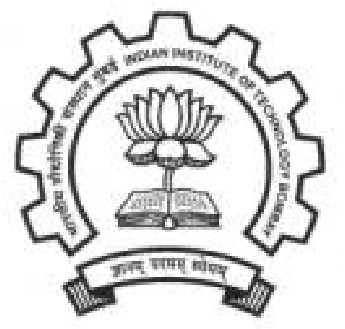}
\end{center}
\end{figure}

Department of Computer Science and Engineering\\
Indian Institute of Technology, Bombay\\
Mumbai
}
\end{center}
\end{titlepage}

\newpage
\chapter*{Acknowledgement\markboth{Acknowledgement}{Acknowledgement}}
\thispagestyle{empty}
\noindent I thank \textbf{Prof. S. Sudarshan} for his inspiration, invaluable guidance and support in making my B.Tech Project an informative and challenging experience for me. His enthusiasm and patience during the last one year have driven me all this while. I would also like to appreciate Prof. Soumen Chakrabarti for his inputs during all the examinations. 

I thank my friends and labmates for bugging me at all the wrong times, but providing support when I needed it the most. A special thanks also to \textbf{Kumar Gaurav Bijay} for his valuable comments and suggestions throughout the project.
\vspace{\baselineskip}
\vspace{\baselineskip}
\vspace{\baselineskip}
\begin{flushright}Nitin Gupta
\end{flushright}

\newpage
\setcounter{page}{1}
\pagenumbering{roman}
\begin{abstract}
In recent years, there has been a lot of interest in the field of keyword querying relational databases. A variety of systems such as DBXplorer \cite{dbxplorer}, Discover \cite{discover} and ObjectRank \cite{objectrank} have been proposed. Another such system is BANKS, which enables data and schema browsing together with keyword-based search for relational databases. It models tuples as nodes in a graph, connected by links induced by foreign key and other relationships. The size of the database graph that BANKS uses is proportional to the sum of the number of nodes and edges in the graph. Systems such as SPIN, which search on Personal Information Networks and use BANKS as the backend, maintain a lot of information about the users' data. Since these systems run on the user workstation which have other demands of memory, such a heavy use of memory is unreasonable and if possible, should be avoided. In order to alleviate this problem, we introduce EMBANKS (acronym for External Memory BANKS), a framework for an optimized disk-based BANKS system. The complexity of this framework poses many questions, some of which we try to answer in this thesis. We demonstrate that the cluster representation proposed in EMBANKS enables in-memory processing of very large database graphs. We also present detailed experiments that show that EMBANKS can significantly reduce database load time and query execution times when compared to the original BANKS algorithms.
\end{abstract}

\newpage
\setcounter{page}{3}
\pagenumbering{roman}
\tableofcontents

\newpage
\listoffigures

\newpage

\setcounter{page}{1}
\pagenumbering{arabic}
\chapter{Introduction}
In recent years, the field of keyword querying relational databases has received great attention. Database search engines have to cope with several challenges such as designing a representation for the database graphs that can efficiently support a diverse range of complex queries over various schema. Given the size and rapid improvement in disk-types, a relatively small database today has millions of tuples. To this end, prior and ongoing research projects such as DBXplorer \cite{dbxplorer}, Discover \cite{discover} and ObjectRank \cite{objectrank} have been proposed. Given a set of query keywords, DBXplorer returns all rows (either from single tables, or by joining tables connected by foreign-key joins) such that the each row contains all keywords. ObjectRank \cite{objectrank}, on the other hand, is an authority-based keyword search engine for databases which returns a group of nodes containing all keywords.

With evolution of different techniques in this area of research, a uniform model has emerged for representing relational databases as a graph with the tuples in the database mapping to nodes and cross references (such as foreign key and other forms of references) between tuples mapping to edges connecting these nodes. Each tuple in the database is modeled as a node in the directed graph and each foreign key-primary key link as an edge between the corresponding tuples. This can be easily extended to other type of connections; for example, it can be extended to include edges corresponding to inclusion dependencies, where the values in the referencing column of the referencing table are contained in the referred column of the referred table but the referred column need not be a key of the referred table. Keywords in a given query activate some nodes. The answer to the query is defined to be a subgraph which connects the activated nodes.

BANKS \cite{banks} (acronym for Browsing ANd Keyword Searching) is a system that enables data and schema browsing together with keyword-based search for relational databases. It models tuples as nodes in a graph, connected by links induced by foreign key and other relationships. Answers to a query are modeled as rooted trees connecting tuples that match individual keywords in the query. Recently, Kacholia et al \cite{banksnew} proposed a bidirectional expansion algorithm for keyword search, an improvement over the original BANKS backward expanding search algorithm. The improvement has been suggested because the original BANKS algorithm performs poorly if some keywords match many nodes, or some node has very large degree. The size of the database graph that BANKS uses is proportional to the sum of the number of nodes and edges in the graph. Systems such as SPIN, which search on Personal Information Networks and use BANKS as the backend, maintain a lot of information about the users' data. Since these systems run on the user workstation which have other demands of memory, such a heavy use of memory is unreasonable and if possible, should be avoided.

In this thesis we introduce EMBANKS (acronym for External Memory BANKS), a framework for an optimized disk-based BANKS system. This framework is intended to support various search models (primarily BANKS) with data structures to facilitate external memory search, and thus alleviate the problem of excessive memory usage. EMBANKS, in a nutshell, proceeds as follows: it first clusters the given input graph based on various parameters and stores the graph thus obtained on disk. The search algorithm is then executed on this smaller graph to return a few answers, which are then expanded and the search algorithm is then re-executed on this expanded graph to get \textit{real} answers. As described herein, EMBANKS, apart from reducing the required memory size, also speeds up the database load time and run time for various queries when compared to the both the existing algorithms. This basis of the framework leads us to a series of questions - which clustering techniques should be used? What should be the size of a single cluster? How will the various weights (for both nodes and edges) in a clustered graph be computed? What is the ideal number of \textit{artificial} answers to generate? We try to provide answers to most of these questions in this thesis.

\section{Problem Identification}
\begin{table}[ht]
\begin{center}
\begin{small}
\begin{tabular}{|cc|c|c|c|c|c|c|c|c|} \hline
 & & \multicolumn{2}{|c|}{DBLP} & \multicolumn{2}{|c|}{IMDB} & \multicolumn{2}{|c|}{IIT Movie} & \multicolumn{2}{|c|}{US Patent} \\ \hline
  &  & Num & Size & Num & Size & Num & Size & Num & Size \\ \hline
Nodes & & 1.77M & & 1.74M & & 4.33K & & 2.23M & \\ \hline
 & nodeType & & 7.09 & & 6.96 & & 0.17 & & 8.94 \\ \hline
 & Prestige & & 7.09 & & 6.96 & & 0.17 & & 8.94 \\ \hline
 & adjacencyOffset & & 7.09 & & 6.96 & & 0.17 & & 8.94 \\ \hline
 & nodeIndeg & & 7.09 & & 6.96 & & 0.17 & & 8.94 \\ \hline
 & nodeOutdeg & & 7.09 & & 6.96 & & 0.17 & & 8.94 \\ \hline
Edges & & 8.49M & & 7.94M & & 21.43K & & 11.88M & \\ \hline
 & AdjacentNode & & 33.99 & & 31.77 & & 0.86 & & 47.54 \\ \hline
 & Weight & & 33.99 & & 31.77 & & 0.86 & & 47.54 \\ \hline
 & Priority & & 33.99 & & 31.77 & & 0.86 & & 47.54 \\ \hline
Total & & & \textbf{137.42} & & \textbf{130.11} & & \textbf{3.43} & & \textbf{187.32} \\ \hline
\end{tabular}
\end{small}
\caption{\label{banks1}The minimum memory requirement (in MBytes) by BANKS}
\end{center}
\end{table}

Table \ref{banks1} lists the minimum memory requirement of BANKS at load time. Though this preprocessing and loading has to be done only once, the numbers show that this minimum requirement is significant even for moderately sized databases. The numbers presented in Figure \ref{banks1} do not include the memory used by pointers and other such objects references which are essential for the execution. It also does not include the memory required for query execution variables such as random strings, priority queues and heaps.

Let us now consider a personal information network. Assuming 3 users, we estimate the number of files on the computer to be roughly half a million, with over five million-edges between them based on proximity and similarity. Given this, the minimum memory requirement of BANKS can be estimated to be $(0.5M \times 4 \times 5) + (5M \times 4 \times 3) = 70MB$!

\section{Organization of the Thesis}
This thesis has three main parts. In the first part, we explore various techniques that have been suggested for other systems that might be applicable for an external memory implementation of the BANKS algorithm. Clustering and multi-level traversal are two such techniques \cite{webgraphs}. Motivated by these systems, we develop EMBANKS through a series of several disk-based optimization techniques for the BANKS algorithm. We summarize the contributions of this part of the thesis in Chapter 3. In the third part of the thesis, we discuss accuracy and efficiency constraints for the EMBANKS framework and suggest solutions to improve the same. We summarize the contributions of this part of the thesis in Chapter 4. In Chapter 5, we present detailed experiments that show that EMBANKS can significantly reduce database load time and query execution times when compared to the original BANKS algorithms. We conclude the thesis with a discussion on the results obtained and possible future work, in Chapter 6.

\chapter{Related Work}
There has been a lot of interest recently in the field of keyword querying relational databases. A variety of systems such as DBXplorer \cite{dbxplorer}, Discover \cite{discover} and ObjectRank \cite{objectrank} have been proposed. An interesting proposal for querying a web graph based on semantics has been proposed in SphereSearch \cite{spheresearch}, which claims to support concept-aware, context-aware, and abstraction-aware search. We begin this chapter with a formal introduction to the graph model used by most of the database search engines today, followed by a detailed explanation of BANKS in Section 2.2, a brief look at the former systems in Section 2.3 and finally conclude in Section 2.4 with the main inspiration for this thesis - the web graph model.

\section{An Introduction to the Graph Model}
The formal graph model used by database search engines can be described as follows:
\begin{packed_item}
\item Vertices: For each tuple $T$ in the database, the graph has a corresponding node $n(T)$. We will speak interchangeably of a tuple and the corresponding node in the graph.
\item Edges: For each pair of tuples $T_1$ and $T_2$ such that there is a foreign key from $T_1$ to $T_2$, the graph contains an edge from $n(T_1)$ to $n(T_2)$ and a back edge from $n(T_2)$ to $n(T_1)$ (this can be extended to handle other types of connections).
\item Edge weights: This weight assignment varies from one technique to another. Weight of a forward link along a foreign key relationship reflects the strength of the proximity relationship between two tuples and is normally set to 1 by default. It may be set to any desired value to reflect the importance of the link (low weights correspond to greater proximity). Let $s(R_1 , R_2)$ be the similarity from relation $R_1$ to relation $R_2$ where $R_1$ is the referencing relation and $R_2$ is the referenced relation. Then the similarity $s(R_1, R_2)$ depends upon the type of the link from relation $R_1$ to relation $R_2$, and is different than the actual edge weights.
\item Node weights: Each node $n$ in the graph is assigned a weight $W(n)$ which depends upon the prestige of the node. In simplest case it can be set to the in-degree of the node.
\end{packed_item}

\section{BANKS}
This section describes BANKS, a system which enables keyword-based search on relational databases, together with data and schema browsing. BANKS enables users to extract information in a simple manner without any knowledge of the schema or any need for writing complex queries. A user can get information by typing a few keywords, following hyperlinks, and interacting with controls on the displayed results. Answers are ranked using a notion of proximity coupled with a notion of prestige of nodes based on inlinks, similar to techniques developed for web search.

The idea of proximity search in databases represented as graphs was also proposed by Goldman et al. They support queries of form find object near object. They restrict results to tuples from one relation near a set of keywords, whereas BANKS permits results to be structured as trees which helps explain how it arrived at an answer. Unlike BANKS, they do not consider node and edge weighting techniques.

\subsection{Backward Expanding Search Model}
The algorithm presented in BANKS \cite{banks} models the database as a directed graph where each tuple is a node of the graph. Foreign-key-primary key links are modeled as directed edges between the corresponding tuples. The edges are directed since the strength of connections between two nodes is not necessarily symmetric. For example, consider the data graph of DBLP, which has a node called conference, connected to a node for each conference, which are then connected to papers published in those conferences. The path through the conference node is a relatively meaningless (compared to an authored-by edge from paper to author). This leads to a natural application of directed edges. The weight of such an edge along a foreign key relationship reflects the strength of the proximity relationship between two tuples. It can be set to any desired value to reflect the importance of the link (small values correspond to greater proximity).

Each answer tree is assigned a relevance score, and answers are presented in decreasing order of that score. The scoring described in the paper involves a combination of relevance clues from nodes and edges. Node weights and edge weights provide two separate measures of relevance. One of the desirables of the algorithm is to control the variation in individual weights so that a few nodes or edges with very large weights do not skew the results excessively.

Finding such a subgraph is a NP complete problem (computation of minimum Steiner trees). This is further complicated by node weight considerations, required to compute the overall relevance of a tree. The \textit{backward expanding search algorithm} described in the paper offers a heuristic algorithm for incrementally computing query results. One of the initial assumptions of the algorithm is that the \textit{graph fits in memory}.

The algorithm begins by looking up tuples containing the search keywords with the help of disk resident indices or symbol tables. Given a set of keywords, for each keyword term $t_i$ , there is corresponding set of nodes, $S_i$ , that are relevant to the keyword. All nodes belonging to any of these sets are marked and the main goal of the algorithm is to find a subgraph connecting these marked nodes. Since just by looking at the subgraph it is not apparent as to what information it conveys, the algorithm tries to identify a node in the graph as a central node that connects all keyword nodes, and also strongly reflects the relationship amongst them.

\begin{figure}[htbp]
\begin{center}
\includegraphics{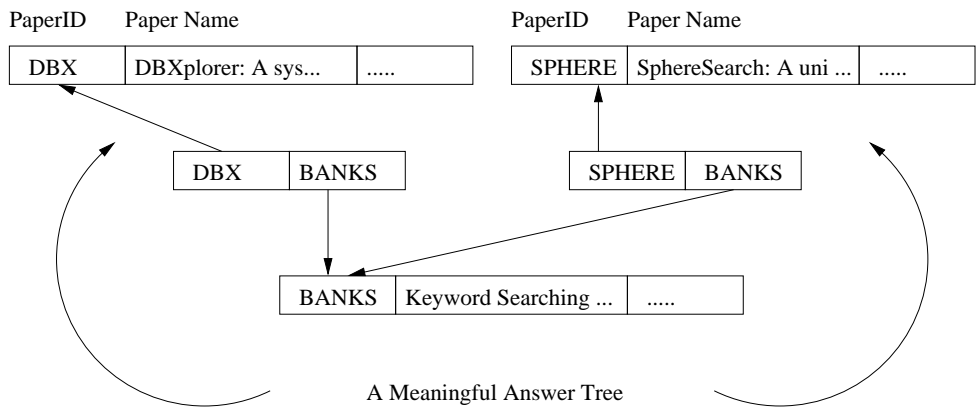}
\caption{\label{answer}Example of an answer returned by BANKS.}
\end{center}
\end{figure}

Let $S = \cup S_i$ . The backward expanding search algorithm concurrently runs $S$ copies of Dijkstra's single source shortest path algorithm, one for each keyword node $n$ in $S$, with $n$ as the source. The copies of the algorithm are run concurrently by creating an iterator interface to the shortest path algorithm, and creating an instance of the iterator for each keyword node. Each copy of the single source shortest path algorithm traverses the graph edges in reverse direction. Basically at each iteration of the algorithm, one of the iterators is picked for further expansion. The iterator picked is the one whose next vertex to be visited has the shortest path to the source vertex of the iterator (the distance measure can be extended to include node weights of the nodes matching keywords). A list of all the vertices visited is maintained for each iterator. Consider a set of iterators containing one iterator from each set $S_i$. If the intersection of their visited vertex lists is non-empty, then each vertex in the intersection defines a tree rooted at the vertex, with a path to at least one node from each set $S_i$. The idea is to find a common vertex from which a forward path exists to at least one node in each set $S_i$ . Such paths define a rooted directed tree with the common vertex as the root and the corresponding keyword nodes as the leaves. The tree thus formed will be a connection tree and root of the tree is the information node.

\subsection{Bidirectional Search Model}

\noindent In brief, backward expanding strategy described above does a best-first search from each node matching a keyword; whenever it finds a node that has been reached from each keyword, it outputs an answer tree. However, Backward expanding search may perform poorly w.r.t. both time and space in case a query keyword matches a very large number of nodes (e.g. if it matches a "metadata node" such as a table or column name in the original relational data), or if it encounters a node with a very large fan-in (e.g. the "paper appeared in conference" relation in DBLP leads to "conference" nodes with large degree). In other words, there are two scenarios in which backward search unnecessarily explores a large number of nodes:

\begin{figure}[htbp]
\begin{center}
\includegraphics{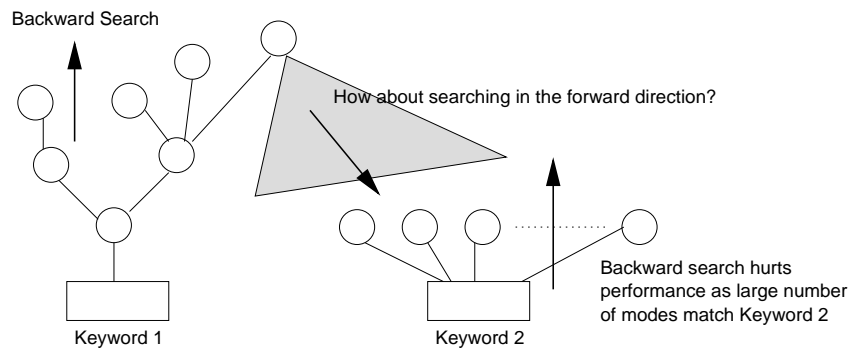}
\caption{Motivation for the new BANKS algorithm.}
\end{center}
\end{figure}

\begin{packed_enum}
\item Since the backward algorithm associates one iterator with every keyword node, if the number of nodes that match the keywords is high (or query contains a frequently occurring term), the algorithm would generate a large number of iterators (e.g. database in the DBLP database, John in   the IMDB database) or if the keyword matches a relation name (which selects all tuples belonging to the relation).
\item The algorithm reaches a node with large fan-in: An iterator might go on to explore a large number of nodes if it hits a node with a very large fan-in (e.g. a department node in a university database which has a large number of student nodes connected to it).
\end{packed_enum}

\noindent Kacholia et al \cite{banksnew} introduce a new search algorithm, which is termed \textit{Bidirectional Search}, for schema-agnostic text search on graphs. The difference between the backward search algorithm and the bidirectional search algorithm is that unlike backward algorithm, which can only explore paths backward from keyword nodes toward the roots of answer trees, the bidirectional algorithm can explore paths forward from nodes that are potential roots of answer trees, toward other keyword nodes.

Another difference is that the bidirectional search runs only one iterator to explore backward paths from the keyword nodes as opposed the multiple iterators for the backward search algorithm. Unlike the backward search algorithm, the iterator is not a shortest path iterator since the nodes cannot be ordered to be expanded solely on the basis of the distance from the origin; the nodes are ordered by the prioritization mechanism described later (activation spread). The benefit of having a single iterator is that the amount of information to be maintained is sharply reduced.

Bidirectional search also maintains another data structure called the outgoing iterator, which expands the nodes in forward direction from potential answer roots. Every node reached by the incoming iterator is treated as a potential answer root. For each root, the outgoing iterator maintains shortest forward paths to each keyword; some of these would have been found earlier by backward search on the incoming iterator, others may be found during forward search on the outgoing iterator.

\textbf{Activation Spread.} One of the most interesting aspects of the bidirectional model is the activation spread. The bidirectional algorithm has just two iterators, so it must prioritize the nodes on some basis for execution. Thus, the paper proposes a novel prioritization scheme based on spreading activation (a kind of Pagerank which decays with distance; also refer to ObjectRank \cite{objectrank}). This technique allows preferential expansion of paths that have less branching, and the same mechanism can be extended to implement other useful features, such as enforcing constraints using edge types to restrict search to specified search paths, or prioritizing certain paths over others.

As described in the paper, the bidirectional search algorithm can work with different ways of defining the initial activation of nodes as well as with different ways of spreading activation. The overall tree score can depend on either the edge score or the node prestige, and both need to be taken into account when defining activation to prioritize search. Nodes matching keywords are initialized with the activation content computed as:

\begin{equation} \label{eqn:prestige}
  \left.
      a_{u,i} = \displaystyle\frac{nodePrestige(u)}{|S_i|}, \forall u \in S_i
  \right.
\end{equation}

\noindent where $S_i$ is the set of nodes that match keyword $t_i$. Thus, if the keyword node has high prestige, that node will have a higher priority for expansion. But if a keyword matches a large number of nodes, the nodes will have a lower priority. The activations from different keywords are computed separately to separate the priority contribution from each keyword. The activation spread from a node is governed by an an attenuation factor $\mu$; each node $v$ spreads a fraction $\mu$ of the received activation to its neighbors, and retains the remaining $1 - \mu$ fraction. As a default the paper uses $\mu = 0.5$.

The activation from keyword $t_i$ is spread to some node $u$ in case of an incoming iterator if there is an edge $u \rightarrow v$. Amongst all such nodes, activation is divided in inverse proportion to the weight of the edge $u \rightarrow v$ (respectively, $v \rightarrow u$). This ensures that the activation priority reflects the path length from $u$ to the keyword node; trees containing nodes that are farther away are likely to have a lower score. For the outgoing iterator, activation from keyword $t_i$ is spread to some node $u$ if there is an edge $v \rightarrow u$, again divided in inverse proportion to the edge weights $v \rightarrow u$. This ensures that nodes that are closer to the potential root get higher activation, since tree scores will be worse if they include nodes that are farther away. When a node $u$ receives activation from a keyword $t_i$ from multiple edges, $a_{u,i}$ is defined as the maximum of the received activations. This reflects the fact that trees are scored by the shortest path from the root to each keyword. Other ways of combining the activation (such as adding them up) could also be used.

\section{Other Systems}
Given the increasing interest in search on relational databases, prior and ongoing research projects such as DBXplorer \cite{dbxplorer}, SphereSearch \cite{spheresearch} and ObjectRank \cite{objectrank} have been proposed. Each of these employs a novel technique for keyword querying the databases. This section discusses the novelty of these systems and draws a critique on each of them.

\subsection{DBXplorer}
DBXplorer is a keyword search utility for relational databases, that has been implemented on top of the Microsoft SQL Server 2000 database server and Microsoft IIS web server. The important features of DBXplorer are:
\begin {packed_item}
\item \textbf {Symbol Table}: It maintains a symbol table that stores mapping between keywords
and database rows where it occurs. The mapping may be at a higher-level granularity if
indices are available, that is, for each keyword we store the columns where
the keyword occurs if an index on the column exists. Some ideas for symbol-table 
compaction are also discussed. 

\item \textbf{Database Representation}: DBXplorer represents the database as an undirected graph where nodes 
correspond to relations (tables) in the database and edges correspond 
to foreign-key links. 

\item \textbf{Answers}: Given a set of query keywords, DBXplorer returns all rows (either
from single tables, or by joining tables connected by foreign-key joins) such 
that each row contains all the keywords.  

\item \textbf{Search Algorithm}: DBXplorer first locates all tables that contain matched keywords by looking into the symbol-table. It then enumerates all such join-trees among the tables such that the set of keywords would be contained in the join-tree. Refer Figure \ref{dbx} where keywords are K1, K2 and K3. Black nodes correspond to tables that contain keywords. Thus, all viable join-trees containing all keywords are identied.

\begin{figure}[htbp]
\begin{center}
\includegraphics{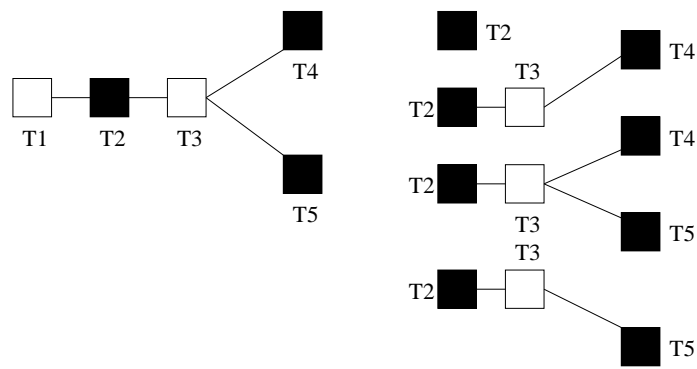}
\caption{\label{dbx}Join Trees in DBXplorer.}
\end{center}
\end{figure}

The joins corresponding to each join-tree is then efficiently carried out; 
information stored in the symbol-table as regards row-id or the
index on the column of the table is exploited while doing so. At an intuitive level, the answer-tree in BANKS can also be seen as a DBXplorer join as edges in the tree are all foreign-key references. Thus, both answer-models are similar.

\item \textbf{Ranking}: For ranking of answers, DBXplorer follows a simple approach. 
DBXplorer ranks answer-rows by the number of joins needed to form them (ties broken arbitrarily). 
\end{packed_item}

\subsection*{Remarks}
\begin{packed_enum}
\item \textbf{Disk-based}: DBXplorer does not represent the database in memory and thus needs less
memory as compared to BANKS. However, there are two issues. Firstly, if a few keywords exist in multiple
tables, enumerating all the interesting join combinations could be costly. Secondly, it is not mentioned
if any attempt is made to utilize the overlapping join-subtrees. If not, this could be quite costly for
metadata keywords or keywords that match a lot of rows. These issues naturally do not arise in 
BANKS.

\item \textbf{Ranking}: DBXplorer's ranking technique is na\"{i}ve. It does not identify important results
from others if they are in the same table or composed of a small number of joins. For example, a 
meaningless answer-row with 2 papers on unrelated topics presented at the same conference is as good as
an answer-row where a paper cites another. Also, there is no way to identify important tuples. So, for the
query 'Pacino' on the imdb.com database, both the legendary 'Al Pacino' of 'Godfather' fame and a 
lesser known supporting actor 'Sal Pacino' have the same rank. This, however, is not an issue with BANKS.
  
\end{packed_enum}

\subsection{ObjectRank}
ObjectRank is an authority-based keyword search engine for databases. The on-the-fly tuning of the system according to the user-specified requirements gives it an edge over other systems. The distinctive features of ObjectRank are: 
\begin {packed_item}

\item \textbf {Database Representation}: Database is viewed as a labeled, directed graph. 
Each node 'v' corresponds to an object where object definitions have to be decided based on database 
schema. Each object may logically correspond to an object in real-life (like the
object-oriented paradigm). For e.g., in the DBLP bibliography schema, there may be nodes corresponding to 
papers, authors, conferences etc. An edge is introduced whenever an object is related to another. 
For example, if paper P1 cites paper P2, an edge is introduced between the two as P1 $\rightarrow$ P2. Or, 
when author A writes paper P, an edge is introduced between the two.

\item \textbf{Answers}: Answers in ObjectRank consist of whole objects. Thus, unlike BANKS or DBXplorer,
no attempt is made to link objects.

\item \textbf{Search Algorithm}: Each node is given global ObjectRank just like PageRank of Google, that is,
based on the random-surfer model. For 
each keyword, there is a keyword-level ObjectRank. So, for each keyword, precompute and save object ranks 
of nodes obtained by setting default PageRanks of only keyword nodes to be non-zero. Object-level PageRank
computation is optimized by defining cut-offs. When a user keys in keywords, these sorted object-level
pageRank lists are fetched and objects in the lists are scored. Objects with the highest score are 
output first followed by others.

\item \textbf{Ranking}: The objects in the lists fetched for each keyword are scored by the formula:

\[
score_{k}(n) = f(GlobalObjectRank(n), KeywordObjectRank_{k}(n))\ \ \ for\ keyword\ k
\]

At run time, their scores are combined :
\[
score_{k1, k2, \dots, km}(n) = score_{k1}(n) * score_{k2}(n) * \dots * score_{km}(n)\ \ \ for\ keywords\ k1, k2, \dots, km 
\]  

\end{packed_item}

\subsubsection{Remarks}
\begin{packed_enum}
\item \textbf{Precomputation}: Keyword-level ObjectRank values must be precomputed and stored
separately for each keyword as calculation of ObjectRank at runtime is computationally expensive. 
However, experiments show that this precomputation is unreasonably expensive. For a database with 3 lakh
nodes and 3 million edges, 74 days of precomputation is necessary on a system with moderate specifications. 
It should be noted that databases of such size are fairly common. For example, the DBLP database has 
about 3 lakh authors, 5 lakh papers and around 15 million edges.

\item \textbf{Object Defintion}: It is unclear as to how objects are defined in the database. As this is
non-trivial, a safe assumption is that objects are manually defined. Defining objects and their edges is 
cumbersome. Also, being a manual task, it is error-prone and not a scalable solution.

\item \textbf{Answer Quality}: As mentioned, ObjectRank does not attempt to link objects as answers. This
 could and does result in answers with high rank that make little sense. For example, if the query is 
'Database Stream Query' in the DBLP database, the user expects to see some paper related to these. 
However, conference nodes where papers with such titles are often published are likely to emerge as 
the top answers.

\end{packed_enum}

\section{A Solution for Web Graphs}
Other systems where the size of the in-memory graph can span over millions of nodes are those which do computations over web repositories. A web repository is a large special-purpose collection of web pages and associated indexes. The Stanford WebBase repository is one such repository. Efficient traversal of huge such web graphs containing several hundred million vertices and a few billion edges is a challenging problem. As a result of the missing schema structure, naive graph representation schemes can significantly increase query execution time and limit the usefulness of web repositories. In \cite{webgraphs}, the authors present a novel way to structure and store web graphs so as to improve the performance of complex queries and computations over web repositories.

In \cite{webgraphs}, the authors propose a novel two-level representation of graphs, called an S-Node representation. In this scheme, a graph is represented in terms of a set of smaller directed graphs, each of which encodes the interconnections within a small subset of pages. A top-level directed graph, consisting of \textit{supernodes} and \textit{superedges}, contains pointers to these lower level graphs. By exploiting empirically observed properties of the graphs to guide the grouping of nodes into supernodes, and using compressed encodings for the lower level directed graphs, S-Node representations provide the following two key advantages:
\begin{enumerate}
\item S-Node representations are highly space-efficient. Significant compression allows large graphs to be completely loaded into reasonable amounts of main memory, speeding up complex graph computations and traversals that require global/bulk access. In addition, this enables the use of simpler main-memory algorithms in place of external memory graph algorithms. Corresponding encoding techniques are described in the next subsection.
\item By representing the graph in terms of smaller directed graphs, the scheme provides a natural way to isolate and locally explore portions of the graph that are relevant to a particular query. The top-level graph serves the role of an index, allowing the relevant lower-level graphs to be quickly located.
\end{enumerate}

\subsection{Graph Structure}
The graph structure as described in \cite{webgraphs} is as follows: let $G$ be the input graph for which a supernode graph needs to be constructed. Let $V(G)$ and $E(G)$ be the vertex set and edge set respectively, of graph $G$. The symbol $p$ refers to a page (or cluster) of the graph, or the vertex if it represents a page (or cluster). Let $P = \{N_1 , N_2 , . . . , N_n\}$ be a partition on the vertex set of $G$. Then the graph structure consists of the following components:

\begin{figure}[htbp]
\begin{center}
\includegraphics{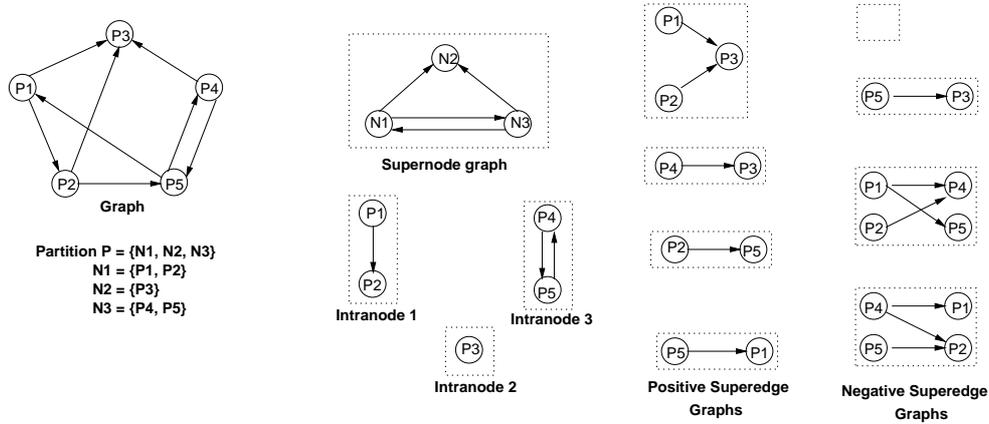}
\caption{Partitioning the graph.}
\end{center}
\end{figure}

\begin{packed_enum}
\item \textbf{Supernode graph.} A supernode graph contains $n$ vertices called supernodes, one for each element of the partition that are linked to each other using directed edges called superedges. Superedges are created based on the following rule: there is a directed superedge $E_{i,j}$ from $N_i$ to $N_j$ if and only if there is at least one node in $N_i$ that points to some node in $N_j$.
\item \textbf{Intranode graph.} Each partition is associated with an intranode graph, which represents all the interconnections between the nodes that belong to $N_i$.
\item \textbf{Positive superedge graph.} A positive superedge graph is a directed bipartite graph that represents all the links that point from nodes in $N_i$ to pages in $N_j$. A positive superedge graph is defined only if there is a corresponding superedge $E_{i,j}$.
\item \textbf{Negative superedge graph.} A negative superedge graph is a directed bipartite graph that represents, among all possible links that point from nodes in $N_i$ to nodes in $N_j$, those that do not exist in the actual graph.
\end{packed_enum}

\noindent Given a partition $P$ on the vertex set of $G$, a S-Node representation of $G$, $S(G, P)$, is a supernode graph that points to set of intranode graphs and a set of positive or negative superedge graphs. Each superedge $E_{i,j}$ points to either the corresponding positive superedge graph or the corresponding negative superedge graph, depending on which of the two superedge graphs have the smaller number of edges. The assumption is that a graph with a smaller number of edges can be encoded more compactly. While this assumption may not always be true for all compression methods, it is nevertheless useful as an approximate heuristic. This is described in more detail in the next section. The choice between positive and negative superedge graphs allows more compact encoding of both dense and sparse interconnections between nodes belonging to two different supernodes.

\subsection{Desiderata}
\begin{packed_enum}
\item Nodes with similar adjacency lists should be grouped together, as much as possible. If such nodes are grouped together in the cluster, compression techniques like reference encoding can be used to achieve significant compression of intranode and superedge graphs.
\item Nodes assigned to a given cluster are connected by edges having high weights or having some lexicographic similarity. These nodes would tend to have a significant percentage of links, and thus might be traversed in a short span of time.
\end{packed_enum}

\subsection{Clustering and other data-mining approaches}
Described below an iterative process to compute a partition on the set of nodes in the graph that satisfy the desirables listed above. Let $P_0 = {N_{01}, N_{02},\dots, N_{0n}}$ be the initial coarse-grained partition. This partition is continuously refined during successive iterations, generating a sequence of partitions $P_1, P_2,\dots, P_f$, i.e., suppose element $N_{ij}$ of partition $P_i = \{N_{i1}, N_{i2},\dots, N_{ik}\}$ is further partitioned into smaller sets $\{A_1, A_2,\dots, A_m\}$ during the $i$ + $1^{st}$ iteration. Then, the partition for the next iteration is $P_{i+1} = \{N_{i1}, N_{i2},\dots, N_{i,j-1}, N_{i,j+1},\dots, N_{ik}\} \cup \{A_1, A_2,\dots, A_m\}$.

The initial partition $P_0$ groups nodes based on spatial locality. In other words, all nodes in the vicinity of another based on edge weights are grouped into one element of the partition. Since the final partition $P_f$ is a refinement of $P_0$, this ensures that $P_f$ satisfies the second desideratum, i.e., all the nodes in a given element of $P_f$ are connected through strong edges.

During every iteration, one of the elements of the existing partitions is picked and further partitioned into smaller pieces. The authors claim to have tried the policy of always splitting the largest (in terms of number of nodes) element, during every iteration. However, this policy as compared with that of picking an element at random from the existing partition did not show much of a difference, i.e., the size and query performance of the S-Node representation produced by either policy was almost identical. Therefore, for the algorithm description, they assumed that an element is chosen at random. Let the element picked during the $i$+$1^{st}$ iteration be $N_{ij}$ of partition $P_i = \{N_{i1}, N_{i2},\dots, N_{ik}\}$. A technique known as Clustered split is then applied for splitting $N_{ij}$. Although clustered split is computationally more expensive, it is used for its performance effects on individual partitions that are smaller in size.

\textbf{Clustered Split.} Clustered Split partitions the nodes in $N_{ij}$ by using a clustering algorithm, such as k-means, to identify groups of nodes with similar adjacency lists. The output of the clustering algorithm is used to split $N_{ij}$ into smaller pieces, one per cluster. Since the authors have used this technique specifically for web graphs, other techniques could also be tried for the kind of graph that BANKS operates upon, like Correlation Clustering or K-Mediods.

To apply clustered split, a supernode graph is first built for the current partition (since it is employed repeatedly in the iteration, the authors state that it may be a good idea maintain the supernode graph throughout the refinement process, incrementally modifying the graph during every iteration). K-means clustering requires that the value of $k$, the number of clusters, be specified apriori. When applying clustered split to element $N_{ij}$, $k$ is initialized with a value equal to the out-degree of $N_{ij}$ in the supernode graph. An upper bound (which is experimentally determined based on time to convergence for run of k-means over smaller graphs) is conceived for the running time of the algorithm, and the execution is aborted if this bound is exceeded. The value of $k$ is then increased by 2 and the process repeated. If k-means repeatedly fails to converge after a fixed number of attempts,  clustered split may be aborted for the current partition and algorithm proceeds to the next iteration.

\textbf{Stopping criterion.} Beginning with the initial partition $P_0$, the partition refinement using is employed using clustered split techniques, until a stopping criterion is satisfied. Ideally, the algorithm should terminate the iteration only if the current partition cannot be refined further, i.e., the clustered split technique is unable to further split any of the elements of the current partition. Since checking for this condition at every iteration is prohibitively expensive, and a stopping criterion that attempts to estimate if the "ideal stopping point" has been reached should be used.

Specifically, the refinement process may be terminated if the algorithm is forced to abort clustered split for $abortmax$ consecutive iterations. Since the element to be split is chosen randomly during every iteration, this criterion can equivalently be stated as follows: iteration is stopped if, in a randomly chosen subset of the partition containing $abortmax$ elements, none of the elements can be further partitioned using clustered split. For the experiments in \cite{webgraphs}, $abortmax$ was fixed to a fraction of the total number of elements in the partition (exact number was 6). A higher value of $abortmax$ increases the accuracy of the estimate, and allows the iteration to run longer, searching for partitions that can be split further. A small value of abortmax may terminate the iteration prematurely, even if further refinements are possible.

\section{Compression}
We now look at the problem of how well the graphs used by BANKS can be compressed. A good compression ratio would allow for more efficient storage and transfer of graphs, and may improve the performance of the algorithm by allowing computation to be performed in faster levels of computer memory hierarchies. Good compression requires using the structural properties of the graph, but as discussed in the previous section, the kind of graphs that BANKS deals with do not belong to any special family of graphs. An example for this motivation is a graph representation that uses 20 bytes per vertex (5 ints) and 12 bytes per edge (3 floats). Total memory requirement is thus $20 \times |V|+12 \times E$. As the database size increases, this number becomes inconvenient. For example, consider a 256MB desktop user has a database stored on his computer with a million nodes and 10 million edges. BANKS would need $20 \times 1m + 12 \times 10m =$ 140MB of memory space. This is the theoretical value and any programming language explodes this further by an factor of at least 2. Thus, there is a need for reducing memory used by BANKS. This section briefly describes algorithms to efficiently compress such graphs, with an assumption that the graph structures have many shared links.

\subsection{Huffman Encoding}
Assuming that the indegrees and outdegrees of nodes follow a Zipfian distribution, i.e., the fraction of pages with indegree $j$ is roughly proportional to $\displaystyle\frac{l}{j^\alpha}$ for some fixed constant $\alpha$, and similarly the fraction of nodes with outdegree $j$ is roughly proportional to $\displaystyle\frac{l}{j^\beta}$ for some fixed constant $\beta$, there is a large variance in degrees. Thus it is natural to consider Huffman-based compression schemes. A simple such scheme goes through the nodes in order and lists the destination of each outedge directed from that node. Each node is assigned a Huffman codeword based on its indegree. To separate the outedges of each node a special stop symbol can be used. This approach achieves significant compression with little complexity; and it can be used in any system that wants to perform efficient computation on the compressed form of the graph.

A lot of variations of the above method have also been suggested. The compression scheme could also be based on the edges directed into each node, whichever is better. In the case where only an isomorphism of the graph needs to be stored, it may help to avoid the stop symbol. Instead, the graph can be treated as an implicit or explicit representation of the outdegree distribution, where the nodes are sorted by outdegree, and list the outedges for each node as before without the stop symbol.

\subsection{Reference Encoding}
\begin{figure}[htbp]
\begin{center}
\includegraphics{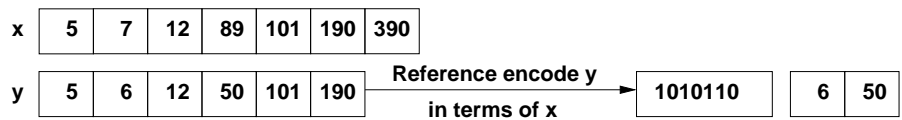}
\caption{\label{refcode}An example of Reference Encoding.}
\end{center}
\end{figure}

Reference encoding \cite{compress} is a graph compression technique that is based on the following idea: if nodes $x$ and $y$ have similar adjacency lists, it is possible to compress the adjacency list of $y$ by representing it in terms of the adjacency list of $x$ (and we say that $x$ is a reference node for $y$). For example, Figure \ref{refcode} illustrates a simple reference encoding scheme. In the figure, the adjacency lists of both $x$ and $y$ contain the entries 5, 12, 101, and 190. The adjacency list of $y$ is encoded using $x$ as a reference node. The encoded representation has two parts. The first part is a bit vector of size equal to the size of the adjacency list of $x$. A bit is set to 1 if and only if the corresponding adjacency list entry is shared between $x$ and $y$ (e.g., third bit is 1 because 12 is shared but second bit is 0 since 7 is not part of $y$'s adjacency list). The second part is a list of all the entries in $y$'s adjacency list that are not present in $x$'s list.

Adler and Mitzenmacher suggest an algorithm such that for a given a graph $G$, for each node $x$, it can decide whether the adjacency list for $x$ is represented as is or in terms of a reference node, and in the latter case, can also identify the particular node that will act as reference. They use the concept of an \textit{affinity} graph $G'$, over which a directed minimum weight spanning tree can be used to generate an optimal (i.e., smallest size) reference encoded representation of $G$. Under appropriate assumptions, the running time of this algorithm is $O(nlogn)$,where $n$ is the number of nodes in the graph. For more details on the algorithm, please refer to \cite{compress}.

As suggested in \cite{webgraphs}, the supernode graph can be encoded using standard adjacency lists in conjunction with a simple Huffman-based compression scheme (each supernode is assigned a Huffman code such that those with high in-degree get smaller codes). The intranode and superedge graphs can also be encoded using the reference encoding scheme described above. In addition, wherever applicable, other easy to decode bit level compression techniques such as run length encoding (RLE) bit vectors or gap encoding adjacency lists can be employed.

These compression techniques are also applicable to the graphs generated by BANKS. However, their impact on efficiency cannot be easily determined. Thus we will not deal with them in this thesis and leave it for future work.

\chapter{EMBANKS}
\section{Introduction}
There are several challenges in designing a representation for BANKS graphs that can efficiently support a diverse range of complex queries over various schema. First, given the size and rapid improvement in disk-types, a relatively small database often has millions of tuples. A representation of these databases must efficiently store and manipulate graphs containing a few million vertices and a few billion edges. If standard data-structures are employed, only a very small portions of the graph will be able to reside in memory. As a result, complex computations and queries become highly memory intensive and time consuming. Second, BANKS graphs do not belong to any special family of graphs (e.g., trees or planar graphs) for which efficient algorithms and storage structures have been proposed in the graph clustering literature. As a result, direct adaptation of compression schemes from these domains is not possible.

In this thesis, we present EMBANKS, a framework that enables in-memory search over data graphs. The dictionary meaning of \textit{embank} is to confine, support, or protect something with an embankment. EMBANKS is a framework that is intended to support various search models (primarily BANKS) with data structures to facilitate external memory search.

\begin{wrapfigure}{r}{3in}
\begin{center}
\includegraphics{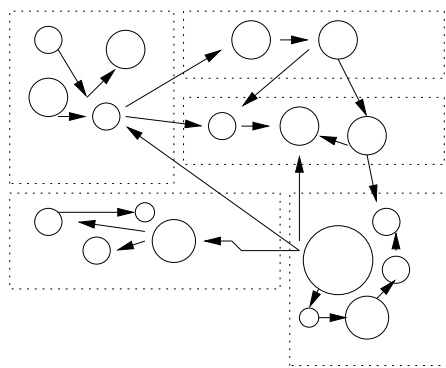}
\caption{Some random clustering of nodes. Notice that area of all the rectangles is equal. Each of them represents one \textit{page} of memory.}
\end{center}
\end{wrapfigure}

EMBANKS, in a nutshell, proceeds as follows: it first clusters the given input graph based on various parameters and stores the graph thus obtained on disk. This clustered graph is organized in such a way that the disk accesses are minimized during runtime. The search algorithm is then executed on this smaller graph to return a lot of \textit{artificial} answers, which are then expanded to form a subgraph of the original graph. The search algorithm is then re-executed on this expanded graph to get \textit{real} answers. This basis of the framework leads us to a series of questions - which clustering techniques should be used? What should be the size of a single cluster? How will the various weights (for both nodes and edges) in a clustered graph be computed? What is the ideal number of \textit{artificial} answers to generate? We try to provide answers to some of these questions through the rest of this section and Section 4.

\textbf{Why cluster?} Clustering comes as a natural solution to the posed problem. It is an important way of exploring graphs, and has been shown to be useful in a wide variety of domains. Informally speaking, clustering is a discovery process that groups a set of nodes such that the intra-cluster similarity is maximized and the inter-cluster similarity in minimized. Clustering graphs provides a means for transforming the system into a smaller, more manageable size thereby by reducing the traversal time considerably.

We define the problem in hand as an optimization problem: Let $G = (V,E)$ be the input graph and $G' = (V', E', C_1 \dots C_k)$ be the graph obtained after clustering. The desiderata is as follows:
\begin{packed_enum}
\item Minimize the time taken (time for processing + number of I/O operations) to obtain and process \textit{artificial} answers. Since $G'$ is in the compressed form, the number of \textit{artificial} answers that should be obtained may be more than those required from the original graph.
\item $G'$ must fit in memory.
\item Minimize the total number of clusters found i.e. $k$. This prohibits the trivial solutions and eliminates trivial solutions such as a clustering with every vertex $v \in V$ in a separate cluster.
\end{packed_enum}

\section{Answer Quality and Accuracy Measurement}
Each answer tree in BANKS (and EMBANKS) has an associated node-score and edge-score:

\begin{equation} \label{eqn:ns}
  \begin{split}
     & nodeScore(answer) = N = weight(root node)+\displaystyle\sum_{l \in leafNodes}weight(l) \\
     & edgeScore(answer) = E = \displaystyle\frac{1}{1+1/\displaystyle\sum_{e\in edges} edgeWeight(e)}
  \end{split}
\end{equation}

These scores are then combined into a final tree score as:

\begin{equation} \label{eqn:ts}
  \left.
  treeScore(answer) = EN^{\lambda}\text{ or }treeScore(answer) = \lambda N+(1-\lambda)E
  \right.
\end{equation}

The results are output from output-heap sorted by their tree scores. By maintaining an upper-bound on the next best answer possible, some results are outputted from the heap before all answers are generated.
 
Based on this scoring model, we define \textit{answer quality} of some answer $a$ as the difference between the score of $a$ and the score of the best answer. A good answer therefore is one whose score falls within some $\epsilon$-margin of the score of the best answer. Answers for a query $q$ having number of nodes and edges less than or equal to the maximum of number of nodes and edges of the top-10 answers obtained from BANKS is henceforth referred to as an \textit{acceptable} answer.

\section{Clustering}
To build a clustered graph that efficiently supports the search and provides some guarantee on the answer quality, we need to identify a partition on the set of nodes in the graph that meet the following requirements:
\begin{packed_item}
\item Each partition should be roughly of the same size, and this size should be a multiple of \textit{page size}. This is crucial to achieving a reduce number of disk I/Os.
\item The partition must be such that given a cluster-level subgraph, we can provide some guarantee on the answer quality for most of the queries, and ensure that the same is not compromised by a significant margin for others (i.e., the score of the answers obtained using EMBANKS is within an $\epsilon$-margin of the scores of answers obtained using BANKS).
\end{packed_item}

\subsection{Comparison Metric}
The aim of graph compression is to get good quality results without the system occupying too much memory and taking too much time. Also, since the search algorithm of EMBANKS is independent of the compression used, it could be made a part of the metric definition. Assuming a human-interactive system, we adopt the following cluster compression metric:

Given two compressed graphs, $G_{C1}$ and $G_{C2}$, $G_{C1}$ is better than $G_{C2}$ if for a sample set of queries, both compute the best answers (say 5 best) with the constraint that the system uses less than some $m$ memory, and $G_{C1}$ results in lesser execution time than $G_{C2}$.

The proposed metric, however, is very coarse in the sense that though it is intuitive, it is hard to use in practice. The trade off between quality of results and the time required for computation imposes limitations to arrive at a theoretically optimal clustering strategy and hence we employ heuristics based on the above comparision metric.

\subsection{Techniques}

The problem of clustering now can be formulated in many ways. Some ways to formulate the problem and their relevance to EMBANKS have been described below.

\subsubsection{Naive (Adjacency) clustering}
The naive clustering technique is mostly generic clustering with certain constraints. The objective is to cluster the nodes having identical adjacency matrices, or nodes which are structurally very similar (similar adjacency lists or similar weights and prestige). It is based on the intuition that the graph size explodes when a keyword is a relation name or an attribute name, and in such cases, quite a few nodes would be similar in terms of structure, though dissimilar in content.

\begin{figure}[htbp]
\begin{center}
\includegraphics{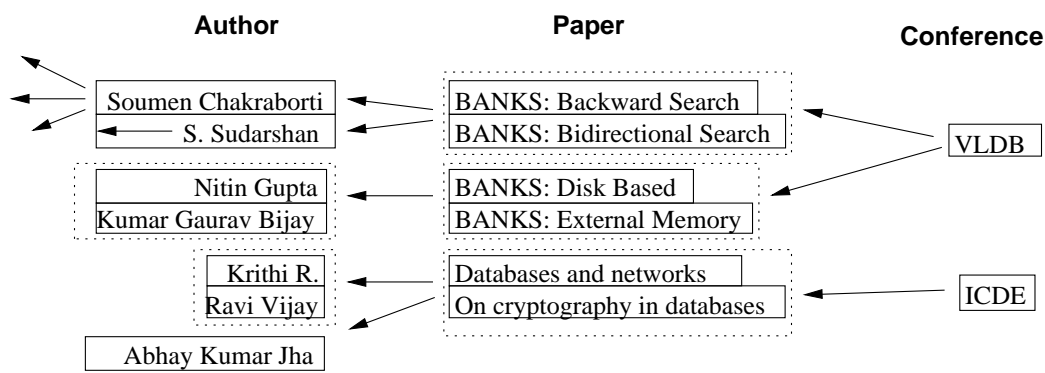}
\caption{An example of Naive (Adjacency) clustering.}
\end{center}
\end{figure}

Though intuitively this approach seems helpful, we realized on experimentation that it fails. This was due to the fact that if the algorithm found a cluster $C$ containing a keyword node $n$, which was connected to another cluster $C'$ through the edge $(m, m') \ni m \in C, m' \in C', m ]ne n$, then even though the compressed graph returned an answer, a \textit{real} answer may not exist.

\subsubsection{Connection clustering}
The failure of the naive-adjacency clustering technique leads to the exploration of connection-based clustering. Let $G = (V,E)$ be the input graph. Let $G' = (V', E', C_1 \dots C_k)$ be the graph obtained after clustering. Then connection clustering is imposes that $\forall i \forall n, m \in C_i \exists n_i \dots n_k \in C_i \ni (n_i, n_{i+1}) \in V$.

We performed experiments with connection clustering and obtained a few answers. For performance results, please refer to Table \ref{clusters}. However, naive connection clustering did not perform very well as there was no way of providing any bound on the answer score. But it did not perform very poorly either as randomization led to an equi-probable distribution of nodes into various clusters.

\textbf{Close-to-1 clustering.} A possible heuristic for the above problem is to minimize the diameter of the clusters to prevent under-weighted answers, i.e., $\forall C_r$, minimize $\max_{i,j \in C_r} sd_{i,j}$, where $sd_{i,j}$ is the shortest distance in $G$ between $v_i$ and $v_j$. This in a way means that we want to minimize the error metric for two nodes inside a cluster.

\textbf{Greedy-Minimum clustering.} Another possible heuristic is to pick up a random node to $n$ represent a cluster $C$, and add to this cluster all nodes adjacent to $n$. If there is more space in the cluster, then iteratively pick a node $n' \in C$ closest to $n$ and add all the nodes adjacent to $n'$.

\subsection{Bounds on Answers}
In order to output answers from the final output heap with confidence, it would be helpful to have bounds on the possible final answers that could be produced by the cluster-level graph. The clusters in an answer-tree in the clustered graph can be divided into 3 types as shown in the figure. Intermediate nodes are of interest to us.

\begin{figure}[!h]\label{fig:bounds}
\begin{center}
\scalebox{0.7}{\includegraphics{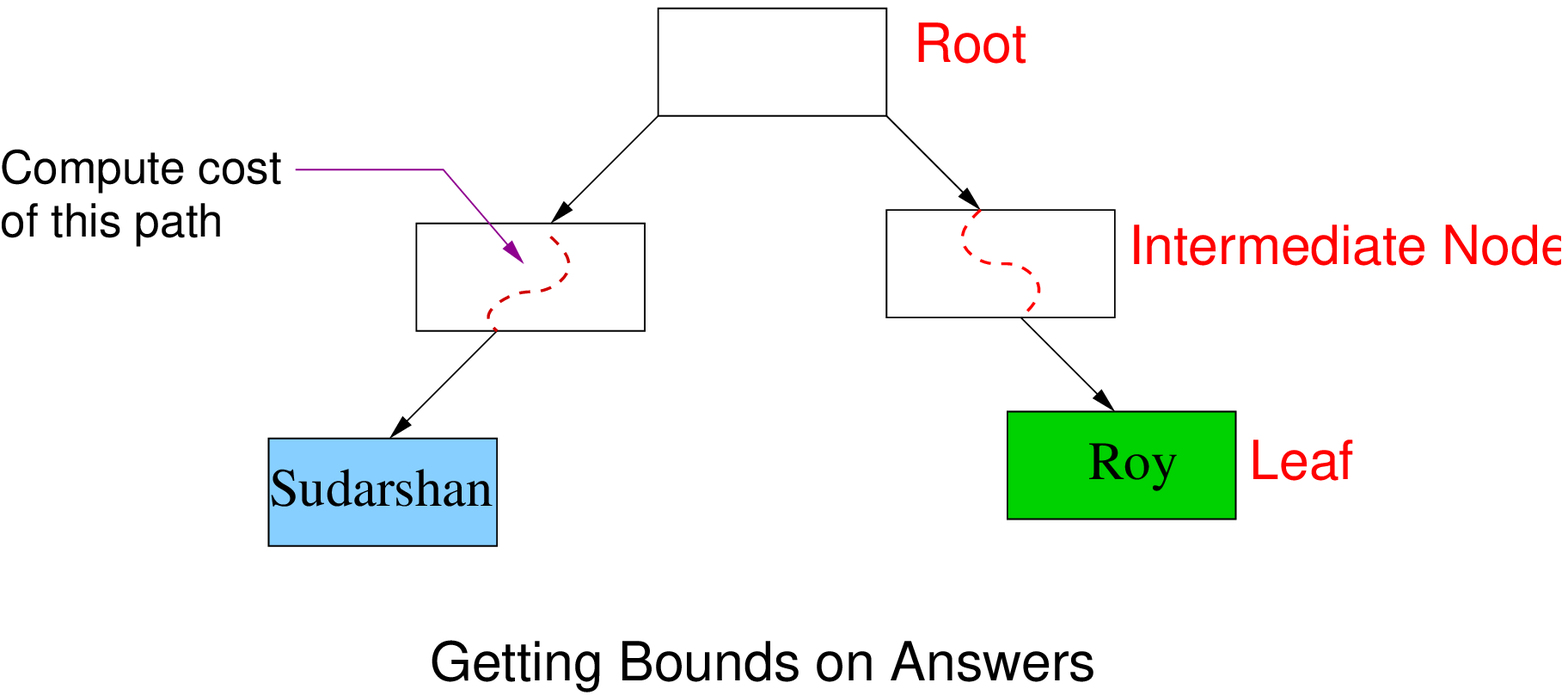} }
\caption{Bounds on answers}
\end{center}
\end{figure}

If, we maintain, for each cluster, the diameter and the cost of the cheapest path from each incoming edge to each outgoing edge (or alternatively just the overall minimum), then for an answer $A = (V,E)$ at the cluster-level, we can say that:

\[ cost(Best Final Answer) \ge \displaystyle\sum_{c \in V'} cost(\text{inedge-outedge pair corresponding to c}) \]

where $V' \subset V$ is the set of clusters that do not contain any keyword and

\[ cost(Best Final Answer) \le \displaystyle\sum_{e \in E} cost(e) + \displaystyle\sum_{c \in V} diameter(c)\]

\subsection{When to get more cluster nodes?}
Search happens in two phases, as discussed. However, for good performance, we need to interleave the two 
phases, as some good final answers might have been missed out for they need not be part of a good answer 
in the cluster-level graph. As a heuristic, we can get new cluster nodes when the score of an answer is lesser than
that of the previous one by a certain factor $\gamma$. So, if 
$score(answer_{i+1}) \le \gamma score(answer_i)$, we can expand more clusters.

\subsection{Experimental Evaluation}
We configured EMBANKS to experiment with bidirectional BANKS on some datasets using the above described clustering algorithms - Naive clustering, Close-to-1 clustering and Greedy-Minimum clustering. Considering the existing BANKS algorithm as the base, we calculated precision on a scale of 10 - the number of answers common between EMBANKS and BANKS, averaged over three random queries per database (DBLP: "sudarshan soumen", "nick roussopoulos christos faloutsos" and "david fernandez parametric"; IMDB: "pierce brosnan james bond", "al pacino diane keaton" and "williams carrey").

\begin{table}[ht]
\begin{center}
\begin{small}
\begin{tabular}{|c|c|c|} \hline
 Clustering & \multicolumn{2}{|c|}{Precision} \\ \hline
 & DBLP & IMDB \\ \hline
 Naive (adjacency) & 0.0 & 0.0 \\ \hline
 Naive (connection) & 4.0 (5, 3, 4)& 0.7 (2, 0, 0)\\ \hline
 Close-to-1 & 4.0 (4, 2, 5) & 4.0 (10, 0, 2)\\ \hline
 Greedy-minimum & 7.0 (5, 9, 7) & 3.7 (9, 2, 0)\\ \hline
\end{tabular}
\end{small}
\caption{\label{clusters}Precision obtained using different clustering techniques}
\end{center}
\end{table}

The numbers in Table \ref{clusters} reflect an \textbf{exact} overlap. Keeping in mind the number of answers produced by BANKS, we manually categorized answers obtained from EMBANKS as acceptable or unacceptable (Refer to Table \ref{relevancy}).

\begin{table}[ht]
\begin{center}
\begin{small}
\begin{tabular}{|c|c|c|} \hline
 Clustering & \multicolumn{2}{|c|}{Precision} \\ \hline
 & DBLP & IMDB \\ \hline
 Naive (adjacency) & 0.0 & 0.0 \\ \hline
 Naive (connection) & 7.0 (7, 4, 10)& 5.7 (10, 1, 6)\\ \hline
 Close-to-1 & 7.0 (7, 4, 10) & 7.0 (10, 3, 8)\\ \hline
 Greedy-minimum & 8.0 (7, 7, 10)& 6.0 (10, 3, 5)\\ \hline
\end{tabular}
\end{small}
\caption{\label{relevancy}Acceptable answers obtained using different clustering techniques}
\end{center}
\end{table}

\section{Weight Adjustment}
Let $G = (V,E)$ be the input graph. Let $G' = (V', E', C_1 \dots C_k)$ be the graph obtained after clustering, and $S = \{(u_1, v_1) \dots (u_s, v_s)\} \subset E \ni \forall i \le s u_s \in C_{r1}, v_s \in C_{r2}$. Then the new edge weights corresponding to $C_{r1,r2}$ and $C_{r2,r1}$ are given by:

\begin{equation} \label{eqn:eweight}
   \begin{split}
                           & \quad \frac{1}{weight(C_{r1},C_{r2})} = \frac{1}{weight(u_1,v_1)} + \dots + \frac{1}{weight(u_s, v_s)} \\
        \text{\textbf{or}} & \quad \frac{1}{weight(C_{r1},C_{r2})} = \frac{1}{s}\biggl(\frac{1}{weight(u_1,v_1)} + \dots + \frac{1}{weight(u_s, v_s)}\biggr) \\
        \text{\textbf{or}} & \quad weight(C_{r1},C_{r2}) = \displaystyle\min weight(u_i,v_i)
   \end{split}
\end{equation}

\noindent The reason for choosing the first two functions is that activation-spread is inversely proportional to the weight of the edge. So, had the nodes not been collapsed, the total activation-spread would have been the inverse sum of the inverse of all contributing edge weights. As this cluster replaces all the nodes, it must receive the whole activation that the combined nodes were receiving. Hence, $weight(C_{r1},C_{r2})$ is given by Equation \ref{eqn:eweight}.

The node prestige of all clusters also need to be defined. Thus $\forall n_i \in C_r$

\begin{equation} \label{eqn:nweight}
   \begin{split}
                           & \quad nodePrestige(C_r) = \displaystyle\sum_{i=1}^{s} nodePrestige(n_i) \\
        \text{\textbf{or}} & \quad nodePrestige(C_r) = \displaystyle\max nodePrestige(n_i) \\
        \text{\textbf{or}} & \quad nodePrestige(C_r) = avg\text{ }nodePrestige(n_i)
   \end{split}
\end{equation}

We further discuss the impact of these choices in the experimental chapter.

\chapter{Implementation}
This chapter explains the implementation of EMBANKS. We first look at the current implementation of BANKS and suggest changes for the same. The second section describes our major contributions.

\section{Current System}
BANKS has been implemented in Java. The current implementation first transforms the given database into an internal graph structure. It builds an inverted keyword-index list which is stored in the database for fast lookup. The database is accessed only twice per se during a query. It is accessed once for an index lookup and then once per answer to determine what text needs to be displayed. The node prestiges, node indegrees, node outdegrees, adjacency lists, edge priorities and edge weights are stored in regular binary files for fast retrieval of bulk data. Table \ref{banksarray} shows how BANKS exactly maintains the data to save memory.

\begin{table}[ht]
\begin{center}
\begin{small}
\begin{tabular}{|c|c|c|c|} \hline
 \textbf{Array name} & \textbf{Type} & \textbf{Size} & \textbf{Function} \\ \hline
 \textit{nodePrestige[]} & float & $|N|$ & $[i]$ is the prestige of node $i$ \\ \hline
 \textit{nodeIndeg[]} & int & $|N|$ & $[i]$ is the indegree of node $i$ \\ \hline
 \textit{nodeOutdeg[]} & int & $|N|$ & $[i]$ is the outdegree of node $i$ \\ \hline
 \textit{adjacencyOffset[]} & int & $|N|$ & $[i]$ is the offset of node $i$'s adjacency list in $adjacentNodes[]$ \\ \hline
 \textit{adjacentNodes[]} & int & $|E|$ & if $e_i=(u,v)$, then $[i]=v$ \\ \hline
 \textit{edgeP[]} & float & $|E|$ & edge priority \\ \hline
 \textit{edgeW[]} & float & $|E|$ & edge weight \\ \hline
\end{tabular}
\end{small}
\caption{\label{banksarray}Various arrays that BANKS creates for graph representation.}
\end{center}
\end{table}

BANKS code has four modules: default, datasource, search and util. The Datasource and Search packages work in sync to process the answers while the default package handles JDBC and HTTP protocols:

\begin{packed_enum}
\item \textbf{Default}: The default package implements the top-level of BANKS - it handles database connections, detects tables and foreign-key references and does precomputation. The package is also responsible for the user interface.

\item \textbf{Datasource}: The datasource package is next in hierarchy. It tskes on the work from the Datasource package and constructs/loads the BANKS graph corresponding to the query.

\item \textbf{Search}: The search package encapsulates the main search algorithm. It begins by parsing the input query. All search algorithms (such as the backward expanding and bidirectional searches) are implemented within this package.

\item \textbf{Util}: The util package is at the lowest level in the hierarchy. It consists of the backbone classes needed to run BANKS. The package implements a 'Tree' class for answer trees and a 'TreeScorer' class for ranking the answers. It also implements Heaps. Java objects use a lot of memory for hashsets, vectors and similar classes. In order to alleviate this problem, BANKS relies on a package known as com.sosnoski \cite{sosnoski} which implements various flavors of hashsets and hashtables like IntHashSet, IntStringHashMap. 
\end{packed_enum}

We observed that BANKS keeps some redundant information in $nodeIndeg[]$ and $nodeOutdeg[]$ for the bidirectional search. Given the above structure, at least one of $nodeIndeg[]$ or $nodeOutdeg[]$ can be eliminated for the backward search algorithm and both for the bidirectional search, saving an additional 5-10\% of memory space.

\subsection{Pruning and other trivial optimizations}
We did some trivial optimizations on the existing graph model in BANKS. The first of these was eliminating $nodeIndeg[]$ and (or) $nodeOutdeg[]$ for the bidirectional (resp. backward) search. Since BANKS introduces reverse edges in the bidirectional search, the indegree and outdegree of the node will always be the same. Thus, we have $\forall i$:
\begin{equation} \label{eqn:biprune}
  \begin{split}
        nodeIndeg[i] = \displaystyle\frac{\text{\textit{adjacencyOffset}}[i+1] - \text{\textit{adjacencyOffset}}[i]}{2} \\
        nodeOutdeg[i] = \displaystyle\frac{\text{\textit{adjacencyOffset}}[i+1] - \text{\textit{adjacencyOffset}}[i]}{2}
  \end{split}
\end{equation}
For backward search algorithm, the indegree and outdegree of a node can be different. Hence, either of the two must be maintained. Therefore, $\forall i$
\begin{equation} \label{eqn:backprune}
  \begin{split}
      nodeOutdeg[i] = \text{\textit{adjacencyOffset}}[i+1] - \text{\textit{adjacencyOffset}}[i] - nodeIndeg[i]
  \end{split}
\end{equation}

We also eliminated a lot of nodes and edges which were meant for transitivity. i.e., if a relation $R$ had only foreign-keys and primary-keys referenced by some other relation, then $R$ is in all probability a transitive relation and has no content of its own. All nodes corresponding to such relations are pruned from the graph.

\section{EMBANKS}
EMBANKS needs to cluster the graph nodes and also expand/read clusters from disks. Searching needs to be done on the graphs twice, which is simply implemented by calling the search function twice.

\subsection{The $Cluster()$ function}
The $Cluster$ function takes as input a graph and maximum cluster size and returns a representation of the clustered graph. This representation of the clustered graph is supported by three arrays: $nodeMapping[]$, $nodeOrder[]$ and $clusterOffset[]$.
\begin{packed_item}
\item $nodeMapping[]: N \rightarrow N'$ is a mapping from a node to a cluster node.
\item $nodeOrder[]: N' \rightarrow N*$ is a set-mapping function from cluster nodes to the set of nodes.
\item \textit{clusterOffset}[]: $N' \rightarrow N$ is the offset for a cluster in the $nodeOrder[]$ array.
\end{packed_item}

Refer to Algorithm \ref{closeto1} which represents close-to-1 clustering. Line 1-5 are merely initialization commands. Of these, the $nodeUsed[]$ array is used to flag nodes that have already been assigned to some cluster. We begin by probing linearly for an unflagged node (line 7). Upon finding such a node, we add it to the cluster queue and traverse it's adjacency list to find another unflagged node, such that the ratio of the weights of forward and backward edge is closest to 1. We repeat this until we have filled up the cluster queue (line 11-30). At line 31, the algorithm assumes that we have formed a cluster. Now lines 33-36 create a mapping and nodeorder for this cluster. $nodeMapping[i] = j$ means that node $i$ has been assigned the cluster $j$. Similarly, $nodeOrder$[\textit{clusterOffset}[$j$] \dots \textit{clusterOffset}[$j$]+$clusterSize$] are the nodes that belong to cluster $j$.
\begin{algorithm}
\caption{Close-to-1 Clustering Algorithm} \label{closeto1}
\begin{algorithmic} [1]
\STATE n=$<$number of nodes$>$
\STATE clusterSize=$<$size of a cluster$>$
\STATE nodeUsed[1 \dots n] = false
\STATE clusterNumber = 0
\STATE orderPointer = 0
\FOR{i = 0 to n}
 \IF{!nodeUsed[i]}
  \STATE queSize = 0
  \STATE queue[queSize++] = i
  \STATE nodeUsed[i] = true
  \WHILE{queSize $<$ clusterSize}
   \STATE minIndex = -1
   \STATE minValue = inf
   \FOR{j = adjacenyOffset[i] to adjacencyOffset[i+1]}
    \IF{!nodeUsed[adjacentNode[j]])}
     \STATE w1 = edgeW[j]
     \STATE w2 = log()
     \IF{w1$>$w2}
      \STATE ratio = w1/w2
      \ELSE \STATE ratio = w2/w1
     \ENDIF
     \IF{ratio $<$ minValue}
      \STATE minValue = ratio
      \STATE minIndex = j
     \ENDIF
    \ENDIF
   \ENDFOR
   \STATE queue[queSize++] = minIndex
  \ENDWHILE
 \ENDIF
 
 \STATE clusterOffset[clusterNumber] = orderPointer
 \FOR{j = 0 to queSize}
  \STATE nodeOrder[orderPointer++] = queue[j]
  \STATE nodeMapping[queue[j]] = clusterNumber
 \ENDFOR
 \STATE clusterNumber++
\ENDFOR

\end{algorithmic}
\end{algorithm}

\subsection{The DiskManager}
Writing and reading clusters to/from disk is a novel idea. As we mentioned in the previous section, ideally each cluster should fit completely on one page and adjacent clusters should be on consecutive pages. Practically, page definitions and adjacent pages cannot be tracked by Java. So, we extend the constraint of a cluster fitting on one page by that of a cluster fitting on a small number of pages, which can be achieved by limiting the number of nodes and edges in a cluster. Having adjacent clusters on adjacent pages is trivially handled by issuing write instructions serially. 

In our implementation, each cluster has a seperate directory, and a cluster and its adjacent cluster are consecutively numbered. Cluster writes are issued in order of their numbering, which tricks the operating-system disk management systems into allocating consecutive pages for the clusters.

We added a class $DiskManager$ to the system which performs all the I/O operations related to clusters. 

The $DiskManager$ has the following important functions:
\begin{packed_item}
\item $readCompressedGraph()$
\item $writeCompressedGraph()$
\item $readCluster()$
\item $writeCluster()$
\end{packed_item}
The $writeCompressedGraph()$ and $writeCluster()$ functions are for preprocessing only. They always occur in sequence and are preceded by a call to the $Cluster()$ function and thus this whole pre-process is time consuming. The $readCompressedGraph()$ function is called once every time the system is restarted and loads the cluster-level graph into memory. $readCluster()$ is called once per cluster to load the clusters contained in the answers obtained in the first phase.

\chapter{Accuracy Constraints}

The numbers in Table \ref{relevancy} show that EMBANKS does not produce as many good answers as BANKS does. The intuitive reason for this is that EMBANKS is not expanding all of the required clusters in the second phase, which means that none of the \textit{artificial} answers contains these clusters. This, in turn, would probably happen either due to a poor scoring model at the cluster level or a weakness in the algorithm in the first phase or the second phase. On deeper exploration, we found a property of the bidirectional BANKS which may lead to this problem. We discuss this in the first section.

\section{The Minimality Syndrome}
Let $S_i$ be the set of nodes containing keyword $k_i$. Then in the directed graph model of BANKS, a response or answer to a keyword query is a minimal rooted directed tree, embedded in the data graph, and containing at least one node from each $S_i$. A resulting tree is an answer tree only if the root of the tree has more than one child. If the root of a tree $T$ has only one child, and all the keywords are present in the non-root nodes, then the tree formed by removing the root node is also present in the result set and has a higher relevance score. This characteristic is known as the \textbf{root-minimality} of directed trees. The next two subsections present two more definitions of minimality for various search algorithms.

\subsection{Minimality of answers}
Let $k_1 \dots k_w$ be the input keywords. Let $A = (V_A, E_A)$ be an answer such that $V_A \in V, E_A \in E$. Then $\nexists A' = (V_A', E_A') \ni$ $A'$ is an answer and $A' \supset A, V' \supset V$. This effectively means that given an answer containing all the keywords, there does not exists any other answer which is a superset of this answers. Each answer thus is a \textit{minimal Steiner tree}.

\begin{figure}[htbp]
\begin{center}
\includegraphics{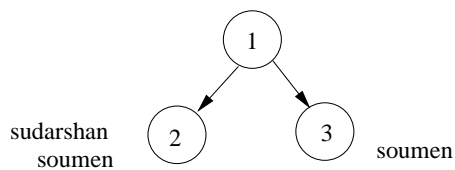}
\caption{\label{steinerfig}An example of a non-minimal answer.}
\end{center}
\end{figure}

Consider the example in Figure \ref{steinerfig}. For the query "sudarshan soumen", the two possible answers are $\{2\}$ and $\{1, 2, 3\}+\{(1,2), (1,3)\}$. However, only the first of these is steiner-minimal. We believe that this characteristic of the answers returned by some search systems (such as XRank) could result in the poor quality of answers produced by EMBANKS.

\subsection{Minimal Intermediate Path}
In order to reduce the amount of information to be maintained, bidirectional BANKS keeps on a single incoming and outgoing iterator for node exploration. Since the algorithm prioritizes nodes using factors other than distance from the keyword node, it is possible that after finding one path from some node $v$ to a keyword $t_i$, it may later find a shorter path; the distance update propagation has to be done each time a shorter path is found. This update propagation results in loss of information with respect to the longer answer - which could have been very useful in the second phase of EMBANKS.

\begin{figure}[htbp]
\begin{center}
\includegraphics{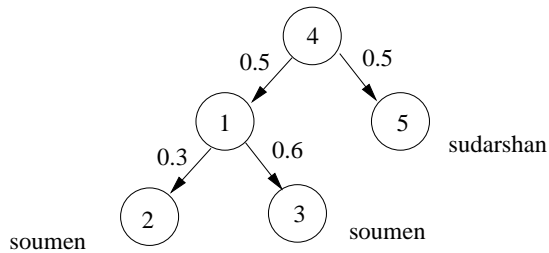}
\caption{\label{minpathfig}A case where the bidirectional search misses answers.}
\end{center}
\end{figure}

Consider the example in Figure \ref{minpathfig}. For the same query as above, i.e., "sudarshan soumen", the graph in consideration should ideally produce two answers. However, $N_1$ stores only the best path to keyword "soumen", i.e., the edge $(1,2)$. This is with the hope that the edge $(1,3)$ will be explored sometime during a forward expansion which may not always happen. This leads to the production of only $\{1,2,4,5\}+\{(1,2), (4,1), (4,5)\}$ and the other answer is lost.

\begin{figure}[htbp]
\begin{center}
\includegraphics{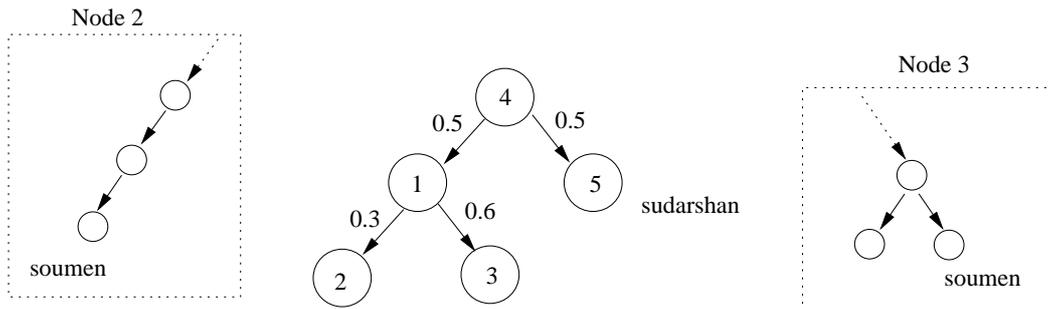}
\caption{\label{pooransfig}A case where the bidirectional search produces poor answers.}
\end{center}
\end{figure}

The seriousness of this problem is reflected in Figure \ref{pooransfig}. The lost answer may be a much better answer after the clusters have been expanded. Bidirectional BANKS on the other hand may not even find this cluster as a part of any of its answer, leading to a serious drop in quality. The backward BANKS algorithm provides us with a solution to this problem since it does not miss any answers. The large number of iterators and the huge task of data maintenance done by the backward search algorithm is now used to produce more diverse answers. This intern leads to the inclusion of a larger number of relevant clusters thereby increasing answer quality.

\section{Utilizing that extra space}
Given a user-specified amount of memory $m$ that BANKS is allowed to use, if the size of the compressed graph plus the size of the clusters expanded after the first phase is less than $m$, EMBANKS expands more clusters that it finds relevant to the given query. We expect that a greater number of nodes in the expanded graph will result in better quality of answers. A naive method to choose these extra clusters is to include all those clusters that have not been selected for expansion after the first phase and contain any of the keywords. If the total size of such clusters exceeds $m$, then we can expand any random subset of these clusters.

\subsection{Introducing extra nodes in the graph}
The second method for utilizing the extra memory space is to introduce more nodes to the graph, other than those obtained from the \textit{artificial} answers. Some such nodes could be the nodes containing the keywords.

Another set of such nodes can be obtained from a parallel clustering of the nodes based on similarity measures other than proximity. A heuristic based algorithm for this kind of clustering begins by sorting keywords based on frequency. Among these keywords, those having frequency greater than some threshold value $f$ are selected. Let these keywords be $k_1,\dots, k_n$. Then using the symbol table or disk resident indices, the tuples (also called nodes) corresponding to these keywords are located and their activation content is initialized to $a_{u,i}$ as discussed in Section 1.3. If a node has multiple keywords, the activation content is simply the sum of activations due to individual keywords. The algorithm then forms clusters in the following manner: It first selects the node having highest activation content and treats it as a cluster $c$. As a result of activation spread from this node, the activations of all nodes adjoining $c$ need to be updated. The next highest node which hasn't been alloted a cluster is then associated with $c$ if and only if the size of the cluster is less than the system page size. In case the node does not fit, a new cluster $c'$ is created and the node with the highest activation is alloted to this cluster. The process is repeated till the coarsest level of the is attained, i.e., all nodes have been associated with some cluster.

\begin{figure}[htbp]
\begin{center}
\includegraphics{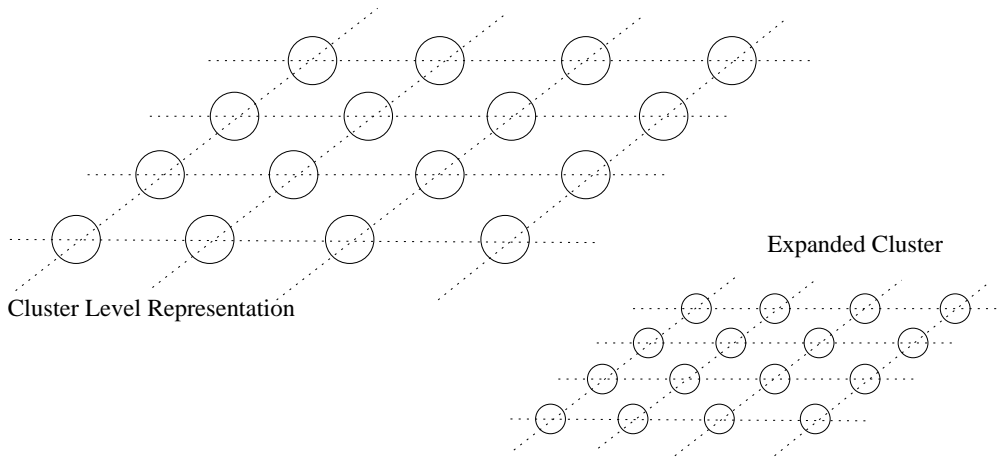}
\caption{\label{granular1fig}The first definition of multigranularity: a two-level graph.}
\end{center}
\end{figure}

We assume that at the end of above iterations, each cluster occupies not more than one page. The following properties should be maintained as the algorithm proceeds: Let $u = group(v_i, v_j)$ be a node obtained by clustering the nodes $v_i$ and $v_j$. Then:

\begin{packed_enum}
\item The activation of the cluster node $u$ is given by $activation(u) = activation(v_i) + activation(v_j)$
\item If the nodes $v_i$ and $v_j$ are derived from the the same relation and have exactly the same adjacency lists (here same means having the same edge weights as well), the adjacency list for $u$ is same as the adjacency list for $v_i$ and the text content of node $u$ is given by $T(u) = T(v_i) \cup T(v_j)$.
\item The cluster must be tightly bound, i.e., $\forall C_i \in C, \forall v \in C_i, \displaystyle\frac{\sum{e(|adj(v)\cap C_i|)}}{|C_i|} \ge \alpha$, where $\alpha$ is some input threshold value.
\end{packed_enum}

An interesting feature of the above approach is that when two nodes are considered for merging, other factors such as ontological similarities can also be integrated without really affecting the algorithm as a whole. This reflects the fact that the algorithm described here may also be applicable not only to BANKS, but to other keyword query engines as well.

\subsection{The second definition of Multigranularity}
\begin{figure}[htbp]
\begin{center}
\includegraphics{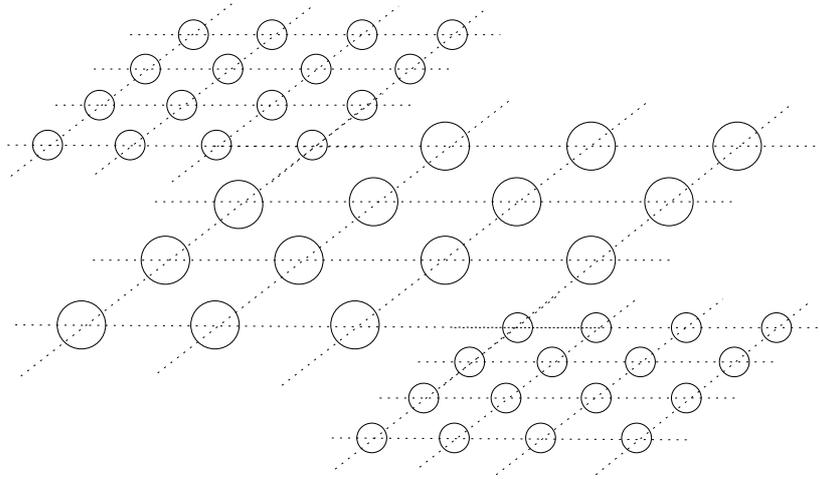}
\caption{\label{granular2fig}The second definition of multigranularity: an interleaved graph.}
\end{center}
\end{figure}

One more way of using the maximum available memory is to expand a few clusters, instead of introducing new clusters into the graph. The graph thus obtained will now have granularity attached to nodes instead of the graph itself. That is, there can be clusters and nodes in the same graph (ref Figure \ref{granular2fig}).

This expansion of clusters can also be done on the fly as and when some answer is generated. This will lead to an iterative expansion of the graph, which will eventually transform into a node-level graph. We expect the answer quality to increase rapidly as the graph becomes finer, since the algorithm can now eliminate a lot of poor answers immediately upon generation.

However, this model also creates a lot of complexities, the most major of them being that the scoring model now needs to be changed. A scoring model based on \textit{indegrees} and \textit{outdegree} of nodes will now not work since nodes and clusters are two separate entities with their own scoring model, and an ad-hoc mixture of the two will not normalize the scores in a desirable manner.

\chapter{Experimental Results}
We conducted several experiments to measure the performance characteristics of EMBANKS using the BANKS algorithm.
\begin{packed_enum}
\item \textbf{Memory Usage:} \textit{How much memory was earlier consumed by BANKS per database and how small have we been able to make it?} Keeping in spirit with the main theme of this thesis, these experiments establish the usefullness of the EMBANKS framework.
\item \textbf{Speed:} \textit{Is there a performance improvement when EMBANKS is used?} We perform experiments to demonstrate that not only does EMBANKS reduce the initial load time by an order of magnitude, but despite the high number of disk accesses, it also outperforms both the BANKS algorithms in terms of time and the number of nodes explored.
\item \textbf{Weight Selection:} \textit{Is harmonic mean a better measure for edge weight or minimum?} We try to answer this and analogous questions with respect to precision of the results.
\end{packed_enum}

\section{Experimental Setup}
BANKS has been implemented in Java v1.5.0 and uses PostgreSQL v7.4.2. The JAVA code connects to the the database using JDBC. Queries are posted through a servlet hosted using the Apache Tomcat web-server v5.0.28.

All the experiments were performed on dual Intel Xeon 3 GHz Processors with EM64T, 4 GB RAM, 2 X 300 GB SATA in RAID-1 hard disk running Debian Linux 2.6.14. The experiments were conducted on the following four datasets:

\begin{packed_enum}

\item \textbf{DBLP}: The DBLP database has 4.28 lakh paper, 2.81 lakh author, 1.11 lakh cites and
9.51 lakh writes tuples. Cites nodes link citations between papers and writes nodes represent the paper-author combinations. Thus, in all, there are 17.71 lakh nodes and 21.24 $\times$ 2 = 42.48 lakh edges.

\item \textbf{US Patents}: The US Patent database has 1.75 lakh category, 4.14 lakh citation,
1.75 lakh coname, 11.44 lakh inventor and 5 lakh patent tuples. Inventor points to patents, patents point to category, coname points to a patent and citations link two patents. Thus, in all, there are 24.08 lakh nodes and 26.47 $\times$ 2 = 52.94 lakh edges.

\item \textbf{IMDB}: The IMDB (Internet Movie Database) has 0.34 lakh cite, 1.04 lakh movie, 6.44 lakh person and 9.58 role tuples. Role tuple maps to a person and a movie. A cite tuple links two movies. Thus, there are 17.40 lakh nodes and 19.84 $\times$ 2 = 39.68 lakh edges.

\item \textbf{IIT Bombay ETD}: The IIT Bombay Thesis database has 30 department, 505 faculty, 11 programs, 1592 students, 1811 thesis and thesis2 tuples. In all, it has 4329 nodes and 5377 $\times$ 2 = 10754 edges.
\end{packed_enum}

As discussed in Section 3.3, we have implemented three clustering ideas: Na\"{i}ve-connection clustering, Close-to-1 clustering and Greedy-minimum clustering. The size of a cluster has been fixed  to 100 for all the experiments. The second phase of BANKS is run after 100 answers have been returned by the first phase.

\section{Memory}
We consider the DBLP database. The DBLP database has 17.71 lakh nodes and 42.48 lakh edges. After clustering, we are left with 17.71 thousand nodes and 13.78 lakh edges. So, BANKS needs 49.9 MB of RAM while EMBANKS needs 18.3 MB of RAM (without keyword to cluster mapping in database) and 11.2 MB (with the mapping in database). A point to note here is that though nodes have substantially reduced, edges have reduced by a small factor only.

\section{Speed}
In this section, we present results from experiments to test if EMBANKS is significantly slower than BANKS. This is expected as the search algorithm runs twice for EMBANKS and the clusters have to be read from the disk. We found that clustering the graph in different ways had little impact on the speed. As the cluster-size is fixed, size of the clustered graph was roughly the same for all clusterings. Also, owing to randomization and connection-based clustering, even the na\"{i}ve technique did not perform poorly in the second phase of the search when the clusters had to be expanded. We compare both the nodes touched and nodes explored to produce the first 10 answers as well as the time taken to generate the first answer. We consider three types of 2 keyword queries : queries where both keywords match many nodes, queries where both keywords match few nodes and queries where a keyword matches many nodes while the other matches few nodes.

\begin{table}[ht]
\begin{center}
\begin{tabular}{|c|c|c|c|c|c|} \hline
\textbf{Type} & \textbf{Query} & \textbf{System} & \textbf{Time (s)} & \textbf{Nodes Touched} & \textbf{Nodes Explored} \\ \hline
(High, High) & Database     & Bidirectional & 4.56 & 274369 & 35441 \\ \cline{3-6}             & Stream       & EMBANKS & 6.29 & 30484 & 9056     \\ \hline
(Low, Low)   & Sudarshan    & Bidirectional & 0.54 & 33524 & 3948 \\ \cline{3-6}
             & Soumen       & EMBANKS & 1.045 & 26578 & 7401     \\ \hline
(Low, High)  & Nick         & Bidirectional & 0.72 & 156131  & 24990 \\ \cline{3-6}
             & XML          & EMBANKS & 1.731 & 44517 & 14072     \\ \hline
\end{tabular}
\caption{Speed: Bidirectional BANKS vs. EMBANKS}
\end{center}
\end{table}

The above table shows that EMBANKS is slower than BANKS. However, the difference is tolerable. In the above experiments, 'Nodes Touched' and 'Nodes Explored' for EMBANKS is the sum of the nodes touched (or explored) in both the clustered and the expanded graph. \textit{It was also observed that caching has a good impact on the speed.  For example, when the query 'Database Stream' was rerun, only 0.482 sec were needed.}

\section{Weights}
We discussed, in Section 3.4, possible choices for Node-prestige and Edge-weight. We did some experiments with two queries on DBLP, $Q_1$: 'Database stream' (both keywords matching many nodes) and $Q_2$: 'Sudarshan Soumen' (both matching few nodes) and found the following result using soft criterion of acceptability as defined earlier.

\begin{table}[ht]
\begin{center}
\begin{small}
\begin{tabular}{|c|c|c|c|} \hline
 \textbf{Edge Weight}   &  \textbf{Node Prestige}& \multicolumn{2}{c|}{\textbf{Precision}} \\ \cline{3-4}                 &               & \textbf{$Q_1$} & \textbf{$Q_2$} \\ \hline
 Min & Sum &\ \ 2\ \ & 7 \\ \hline
 Min & Max & 0 & 7 \\ \hline
 Min & Avg & 0 & 7 \\ \hline
 Harmonic Mean & Sum & 2 & 6 \\ \hline
 Harmonic Mean & Max & 0 & 7 \\ \hline
 Harmonic Mean & Avg & 0 & 7 \\ \hline
 Inverse Sum & Sum & 2 & 8 \\ \hline
 Inverse Sum & Max & 0 & 7 \\ \hline
 Inverse Sum & Avg & 0 & 7 \\ \hline
\end{tabular}
\end{small}
\caption{\label{weights}Number of relevant answers obtained with EMBANKS.}
\end{center}
\end{table}

Thus, experimentally for the query-set discussed, we find that there is high correlation between
the the choice of node-prestige function and the precision. We found that the combination of inverse sum for edge-weight (parallel resistance) and sum for node-prestige works best.

\paragraph{Remark:}
It is worth noting that the performance of EMBANKS has been bad on the `database stream' case. This prompted us to further analyze this case. We chose few similar queries and experimented as shown in the Table \ref{tab:remark}.

In the following table, `origin size' denotes the number of tuples in the database matching the corresponding keywords. Setting limits on the number of answers produced in the first phase will impact the quality of final answers. So we experimented with two different limits on the number of answers produced in the first phase: 100 and 400.
\begin{table}[ht]
\begin{center}
\begin{small}
\begin{tabular}{|c|c|c|c|} \hline
 \textbf{Query} & \textbf{Origin Size}   & \multicolumn{2}{c|}{\textbf{Precision}} \\ \cline{3-4}
                &                        & \textbf{limit=100} & \textbf{limit=400} \\ \hline
 database stream  &  (7595, 411)   &   2  & 5  \\ \hline
 john xml         &  (3218, 1450)  &   8  & 8  \\ \hline
 time concurrency &  (12734, 1194) &   6  & 10 \\ \hline
 xml query        &  (1450, 3236)  &   10 & 10 \\ \hline
\end{tabular}
\end{small}
\caption{\label{tab:remark} Experiments with keywords matching many nodes}
\end{center}
\end{table}

For most queries matching many keywords, there are many good answers and we find a chunk of them. For 'database stream' however, there are only a few top-quality answers (12) and we find few of them only.

Getting more answers helps our case as more clusters are introduced. That this idea failed for other
type of keywords but works for keywords with big origin size. This indicates that we could use this as a heuristic for adding more clusters.

Furthermore, experiments show that whenever we start missing answers, there is a stark difference in the tree-scores of consecutive answer-trees. This can thus be an indicator of when to fetch new clusters.

\chapter{Conclusion}

\noindent\textbf{Conclusion.} As the amount of information stored in databases increases, so does the need for efficient search algorithms. Keyword search enables information discovery without requiring from the user to know the schema of the database, SQL or QBE-like interface, and the roles of various entities and terms used in the query. Given the current systems, an increase in database size requires an increase in memory, which is unacceptable for simple workstations. The first part of the thesis identified this problem in BANKS and looked into other similar systems facing the same problem. We also looked at techniques suggested for an analogous system \cite{webgraphs}.

The second part of the thesis introduced EMBANKS, a framework for an optimized disk-based search system, which is intended to alleviate the system of the aforementioned problem. EMBANKS has been developed with a vision to support various search models (primarily BANKS) with data structures to facilitate external memory search. As was described, EMBANKS, apart from reducing the required memory size, also sped up the database load time and run time for various queries when compared to the both the existing algorithms.

However, as reflected by the numbers in Table \ref{relevancy}, EMBANKS did not produce as many good answers as BANKS did. A possible reason for this was that EMBANKS did not expand all of the required clusters in the second phase, which meant that none of the \textit{artificial} answers contained these clusters. This, in turn, happened either due to a poor scoring model at the cluster level or a weakness in the algorithm in the first phase or the second phase. The third part of the thesis discussed properties of the bidirectional BANKS which may have led to this problem, and measures to rectify the same.

We have demonstrated that the cluster representation proposed in EMBANKS enables in-memory processing of very large database graphs. We have also established through detailed experimentation that EMBANKS can significantly reduce database load time and query execution times when compared to the original BANKS algorithms.

\vspace{\baselineskip}

\begin{figure}[h]
\begin{center}
\includegraphics{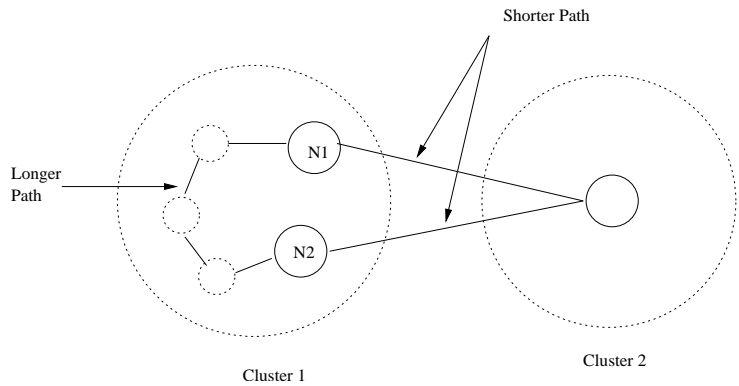}
\caption{\label{shortpath}A shorter path through an external cluster.}
\end{center}
\end{figure}

\textbf{Future Work.} One area that still needs to be worked on relates to compression of clusters. Compression is crucial for global/bulk access computations and mining tasks and thus may help reduce a lot of disk-accesses and lead to increasingly large cluster sizes. Questions such as how big is a cluster representation, or equivalently, how big a search graph can we represent in a given amount of main memory using some clustering scheme are still unanswered. 

A lot of good answers that are generated by BANKS are not found using EMBANKS. This is mostly due to some missing clusters which were essential for an answer, but were not included in the set of clusters that are expanded. An example for this is the Naive Clustering based system, where if two keyword nodes are in the same cluster, but are not connected through a path in that cluster (or are connected through a long path in the general case), then an answer will not be found. However, these nodes could actually be connected through a short path if some neighboring cluster had also been expanded (Refer to Figure \ref{shortpath}). Thus, this inclusion of more clusters based on a given set of clusters provides a window to an area of future work.

Another potential area of work is adaptive query processing and it's application to the introduction of extra nodes. That is, given a keyword query, the algorithm should be able to determine (based on statistics computed over time) which clusters may be useful for that query. The algorithm could then also determine how many answers does it need to find at the cluster-level for expansion.

\nocite{banks,banksnew,objectrank,spheresearch,discover,externalmem,webgraphs,roadmap,filestruct,compress}

\bibliography{report}
\bibliographystyle{alpha}

\end{document}